# Breakdown of Ohm's Law by Disorders in Low-Dimensional Transistors

Chang Niu, Adam Charnas, Jian-Yu Lin, Linjia Long, Zehao Lin, Zhuocheng Zhang and Peide D. Ye*

*Elmore Family School of Electrical and Computer Engineering and Birck Nanotechnology Center, Purdue University, West Lafayette, IN 47907, United States.*

*Correspondence and requests for materials should be addressed to P. D. Y. (yep@purdue.edu)

**Abstract**

Ohm's law provides a fundamental framework for understanding charge transport in conductors and underpins the concept of electrical scaling that has enabled the continuous advancement of modern CMOS technologies. As transistors are scaled to even smaller dimensions, device channels inevitably enter low-dimensional regimes to achieve higher performance. Low-dimensional materials such as atomically thin oxide semiconductors, 2D van der Waals semiconductors, and 1D carbon nanotubes, have thus emerged as key candidates for extending Moore's law. Here, we reveal the fundamental distinction between three-dimensional and low-dimensional conductors arising from disorder-induced electron localization, which leads to the breakdown of Ohm's law and lateral linear scaling. We develop a quantitative model that captures the role of the disordered region, a unique characteristic intrinsically to low-dimensional transistors. Furthermore, the disorder-induced localization framework consistently explains experimental observations in atomically thin $In_2O_3$ field-effect transistors across variations in channel length, temperature, thickness, and post-annealing conditions. This work establishes a unified physical picture for understanding and optimizing disorder-driven electronic transport in low-dimensional transistors.

Ohm's law, which relates a current flow to an applied low voltage through a constant conductance, forms the foundation of classical electronics and enables the operation of all semiconductor devices.[1] However, as transistors are scaled toward atomic dimensions, the assumptions of ohmic transport, diffusive motion of carriers and size-independent conductivity begin to fail. In this regime, disorder, quantum interference, and finite-size effects dominate electron dynamics, giving rise to localization phenomena that fundamentally alter current flow. Understanding and controlling these effects is essential not only for advancing quantum and low-dimensional physics but also for extending transistor performance beyond conventional CMOS scaling limits.[2–4]

In low-dimensional systems, disorder has a dramatically amplified impact on charge transport. Structural imperfections, surface roughness, interface potential fluctuations, and intrinsic material disorder can confine electron wavefunctions, suppress mobility, and even cause a breakdown of Ohm's law. Theoretical frameworks predict that all two-dimensional (2D) systems are inherently localized at zero temperature[5,6], yet the disorder-induced localization effect on field-effect transistors (FETs) is rarely investigated and modeled.[7] Building on the fundamental scaling theory of localization, we establish the connection between disorder strength and the measured current-voltage characteristics in thin-film transistors, providing a unified explanation for the experimentally observed breakdown of Ohm's law. We emphasize the importance of disorder-induced localization effects in FETs formed by atomically thin atomic-layer-deposited (ALD) $In_2O_3$ channel, which is a promising candidate for back-end-of-line (BEOL) monolithic integration due to its excellent device performance.[8–14] The ALD-grown $In_2O_3$ material system provides an ideal platform for investigating electron localization in two dimensions, owing to its natural surface accumulation that maintains conductivity even at sub-nanometer thicknesses.[8,15,16] By systematically controlling channel thickness, annealing temperature, and gate-tuned carrier density, we map the transition from diffusive to localized transport directly and extract the scaling function $\beta(\sigma)$ predicted by theory. The results reveal an exponential suppression of conductance with channel length, a gate- and temperature-tunable localization length, and a universal scaling behavior. Beyond its fundamental significance, this work establishes disorder as a key design parameter in 2D transistor technologies and provides a framework for engineering reliable, high-performance semiconductor devices in the era of atomic-scale electronics.

**Degree of Disorder and Electron Localization in Low Dimensions**

Electron transport in solids can be viewed as the propagation of Bloch waves, where $k_F$ defines the wavevector (and thus the wavelength $\lambda_F$) of the electronic state at the Fermi level. When the electron wavelength exceeds its mean free path $l$, the wave is scattered before completing a full oscillation, causing destructive quantum interference among multiple scattering paths that confines the electron to a localized region in real space as illustrated in **Fig. 1a**. The degree of disorder can be quantitively described by the dimensionless parameter $k_F l$, which represents the ratio between the electron mean free path and its wavelength.[17–19] When $k_F l \sim 1$, the electron wavelength becomes comparable to its mean free path, making the Ioffe-Regel limit that defines the critical transition from metallic to insulating behavior. For $k_F l \gg 1$, electron phase coherence is weakly disturbed and classical diffusive transport (Ohm's law) remains valid. In contrast, for $k_F l \ll 1$, the electron wavelength exceeds the average scattering length, leading to spatial confinement of carriers, a phenomenon known as Anderson localization, which results in the breakdown of Ohm's law (**Fig. 1b**). Moreover, $k_F l$ depends strongly on dimensionality, and in this work, we emphasize the crucial role of disorder in determining electron transport in 2D transistors. We derive $k_F l$ for one-, two-, and three-dimensional conductors as follows (see details from **Note S1**):

$$k_F l = \frac{\hbar\mu}{e}\left(\frac{6\pi^2 n_{3D}}{g_s g_v}\right)^{\frac{2}{3}}, \quad \frac{\hbar\mu}{e}\left(\frac{4\pi n_{2D}}{g_s g_v}\right) \text{ or } \frac{4\pi\hbar}{g_s g_v e^2}\sigma_\square, \quad \frac{\hbar\mu}{e}\left(\frac{\pi n_{1D}}{g_s g_v}\right)^2$$

where $\hbar$ is the reduced Planck constant, $e$ is the elementary charge, $g_s$ and $g_v$ are the spin and valley degeneracy factors, respectively, $n$ is the carrier density, μ is the carrier mobility, and $\sigma_\square$ is the sheet conductance in 2D. **Fig. S1** provides a schematic illustration of $k_F l$ scaling across different dimensions. The carrier-density dependence of the disorder parameter scaling rapidly in lower dimensions and saturates in three dimensions, indicating that disorder-induced transport effects are much more pronounced in low-dimensional systems and become more critical in atomic-scale transistors. More thoroughly investigations are needed.

It is important to note that, in two-dimensional systems, $k_F l$ depends solely on the sheet conductance, providing strong theoretical justification for the critical role of disorder-induced localization in electronic transistors. Since transistor operation inherently requires switching over several orders of magnitude in conductance, the channel inevitably traverses the transition between the localized and diffusive transport regimes. We apply the fundamental concept of electron localization to describe the disorder-induced transport phenomenon in 2D transistors, where a disordered region is introduced to model the current flow. In the presence of localization, conductivity scaling deviates from Ohm's law and instead follows an exponential decay with channel length. This correction to the drain current can be expressed as (**Fig. 1c**):

$$I_D = \sigma_{disorder} V_{DS} = \sigma_0(V_{GS}) exp\left(-\left(\frac{L}{\xi(V_{GS})}\right)^\gamma\right) V_{DS}$$

$$\sigma_0(V_{GS}) = \mu C_{ox}\frac{W}{L}(V_{GS} - V_{th}); \quad \xi(V_{GS}) = \xi_0 exp(\alpha V_{GS})$$

where $\xi$ is the localization length, $\alpha$ is the gate-controlled delocalization coefficient, $\gamma$ is the localization scaling exponent, $\gamma = 1$ indicates strong localization. **Fig. 1d** presents a conceptual comparison of transfer characteristics and transconductance ($g_m$) between a conventional 3D FET (Si-based) and a disordered 2D FET (oxide semiconductors or 2D van der Waals materials). A disorder-dominated region emerges between the subthreshold and linear regimes. At high carrier density (large gate bias), the disorder potential is screened by free carriers, and transport follows classical diffusion consistent with Ohm's law. At low carrier densities, however, the conductivity exhibit corrections arising from disorder-induced localization. In the subthreshold regime, the current is governed by thermally activated emission. The $g_m$ behavior reflects the contrasting switching dynamics between two conditions, where a localization-induced $g_m$ peak is commonly observed in disordered oxide semiconductors[20–22] and 2D materials[23]. **Fig. 1e** and **Fig. S2** illustrate the evolution of the normalized current transfer characteristics as a function of the parameters $\alpha$, $\xi$, and $L_{ch}$. The channel-length-dependent behavior arises solely from the localization effect, reflecting the exponential suppression of conductance with increasing channel length.

**Electron Localization in In$_2$O$_3$ FETs**

We employ atomically thin, ALD-grown In$_2$O$_3$ FETs to systematically investigate disorder-induced localization in two-dimensional semiconducting channels experimentally for the first time. **Fig. 2a** shows a schematic of the device structure, consisting of In$_2$O$_3$ layers with thicknesses ranging from 0.8 to 3 nm and a 5 nm HfO$_2$ gate dielectric. Devices with different channel lengths were fabricated to enable a systematic study of localization effects across various carrier densities and length scales.

The surface-accumulation-induced negative Schottky barrier provides excellent electrical contact to the In$_2$O$_3$ channel.[24] **Fig. 2b** and **2c** show the transfer characteristics of 0.8 nm thick In$_2$O$_3$ FETs with channel lengths ranging from 0.05 to 2 μm, plotted on linear and logarithmic scales, respectively. The devices exhibit a pronounced channel-length-dependent $V_{th}$ shift originating from *electron localization*. This shift persists across the entire range of channel lengths and remains observable even in extremely large samples (on the millimeter scale, see **Fig. S3**), in great contrast to the conventional short-channel effect, where the $V_{th}$ roll-off occurs only at very small channel lengths[25]. A similar channel-length-dependent shift in $V_{th}$ has also been observed in 2D-material-based FETs,[23] which can also be explained by the same localization theory. The normalized conductance, defined as $\sigma_{normal} = I_D \cdot L_{ch}/(V_{DS} \cdot W_{ch})$, and shown in **Fig. 2d**, clearly deviates from Ohm's law, under which no channel-length dependence is expected. Instead, $\sigma_{normal}$ decreases by several orders of magnitude as the channel length increases from 0.2 to 2.0 µm at fixed carrier densities (gate biases). Plotting the normalized conductance on a logarithmic scale as a function of channel length at constant $V_{GS}$ (**Fig. 2e**) reveals a linear relationship, indicating an exponential decay of conductivity with increasing length described by $\sigma_{normal} = \sigma_0(V_{GS}) exp\left(-\frac{L_{ch}}{\xi(V_{GS})}\right)$, which is the signature of disorder-induced strong electron localization. The localization length $\xi$ quantifies the characteristic spatial extent over which an electron's wavefunction remains coherent before becoming localized in real space. As shown in **Fig. 2f**, the gate-dependent localization length extracted from the slopes in **Fig. 2e** increases with $V_{GS}$, reflecting the screening of disorder at higher carrier densities and demonstrating the gate tunability of disorder in 2D transistors. These results provide the compelling evidence for the first experimental observation of electron localization in atomically thin In$_2$O$_3$ oxide semiconductors.

**Diffusion to Localization Transition**

The role of disorder differs fundamentally across electronic dimensionalities.[26] In 3D conductors, electronic transport typically remains diffusive even under moderate disorder, as the disorder parameter $k_F l$ often exceeds the critical threshold for localization.

In contrast, 2D and 1D systems can easily undergo a transition from diffusive to localized transport, driven by variations in the degree of disorder. In this session, we examine how this diffusion-localization transition is modulated by gate bias (carrier density) and temperature.

**Fig. 3a** shows the channel-length-dependent transfer characteristics of $In_2O_3$ FETs with a 2.2 nm channel thickness after post-annealing at 290 °C in an $O_2$ environment. The increased thickness and high-temperature annealing reduce surface roughness, mitigate quantum confinement effects[27], and partially crystallize the $In_2O_3$ film[12], thereby enhancing structural order. Despite these improvements, disorder-induced electron localization remains evident, as indicated by the exponential decay of normalized conductance with channel length (**Fig. 3b**). A clear transition is observed at $V_{GS} = 1.25\ V$ (marked by the grey dashed line), below which the system exhibits strong localization and above which the conductivity becomes independent of channel length, signifying diffusive transport. This crossover is further illustrated in **Fig. 3c**, where the normalized conductivity decreases exponentially at low carrier densities but remains nearly constant at high carrier densities, returning to ohmic behavior. It is worth noting that the observed localization-diffusion transition occurs at a normalized conductance of approximately 38.7 µS/ □, corresponding to the quantum conductance $\frac{e^2}{h}$. This experimental value coincides with the Ioffe-Regel limit,[28,29] which defines the critical point between diffusive and localized transport in 2D, where $\sigma_{critical} = \frac{e^2}{h}$, $k_F l = \frac{4\pi\hbar}{g_s g_v e^2} \sigma_{critical} = 1$, assuming a spin-valley degeneracy of $g_s g_v = 2$. The correspondence between the experimental transition conductance and the theoretical Ioffe-Regel limit provides strong evidence for the critical role of disorder-induced electron localization in 2D transistors. **Fig. 3d** plots the extracted gate-dependent localization length, where a longer localization length indicates weaker localization and a smaller conductivity correction. The convergence of localization length at $V_{GS} = 1.25\ V$ marks the transition from the strongly localized regime to the weak-localization or diffusive transport region, as tuned by the gate voltage. **Fig. 3e** shows the gate-dependent localization length for $In_2O_3$ channels with different thicknesses and annealing temperatures (see **Fig. S4-S7**), each representing a distinct degree of disorder extracted using the same analysis method. The disorder characteristics in these channels can be described by the relation $\xi(V_{GS}) = \xi_0 exp(\alpha V_{GS})$, where $\xi_0$ reflects the intrinsic degree of disorder in the channel, and $\alpha$ quantifies the efficiency of gate-induced modulation of the localization length. **Fig. 3f** summarizes the variation of $\alpha$ with film thickness and annealing temperature, indicating progressive ordering of the system with increased thickness and thermal treatment.

**Fig. 4a** shows the temperature-dependent transfer characteristics of $In_2O_3$ FETs with a 2.2 nm channel thickness and no post-annealing treatment. The negligible hysteresis confirms reliable device operation and a low density of interfacial traps. By analyzing the channel-length dependence at various temperatures (see **Fig. S8**), the temperature dependent disorder-induced localization can be quantitatively extracted. **Fig. 4b** presents the normalized conductivity scaling at a fixed $V_{GS}$ for different temperatures, revealing a temperature-driven transition from localized to diffusive transport. The gate-dependent localization length extracted at each temperature is shown in **Fig. 4c**; both $\xi_0$ and $\alpha$ increase with temperature, indicating a reduction in the effective degree of disorder. Furthermore, **Fig. 4d** displays the temperature dependence of $\alpha$ on a logarithmic scale, where the linear trend predicts the disappearance of gate tunability ($\alpha \to 0$) as $T \to 0\ K$. Theoretically, as $T \to 0\ K$, conduction and electrostatic tunability are expected to vanish in the 2D limit, as all electronic states become spatially localized according to Anderson localization theory. At room temperature, however, diffusive conduction can still occur under weak disorders, while progressive increases in disorder reveal localization effects across different transport regimes. As the active channel thickness approaches the atomic scale, the ability to model and control disorder becomes critically important for both fundamental understanding and device optimization.

**Scaling Theory of Electron Localization**

Finally, we employ atomically thin $In_2O_3$ FETs to experimentally extract the scaling theory of electron localization predicted by Abrahams *et al.* in 1979.[5] While the scaling theory of localization has been extensively examined through temperature- and disorder-dependent transport studies,[30,31] this work represents the first direct experimental realization of the full scaling function $\beta = \frac{dln\sigma}{dlnL}$ (where $\sigma$ is the conductivity and $L$ is the size of the system) obtained from length-dependent conductance measurements in a controlled 2D electronic platform. In our devices, the gate provides an additional degree of freedom that allows continuous tuning of the carrier density through electrostatic control, enabling the full range of conductivity to be accessed within a single set of transistors sharing identical structural disorder. **Fig. 5** presents the experimentally extracted scaling function in two dimensions. The function $\beta$ is calculated from the channel-length-dependent conductivity variation at fixed carrier densities ($V_{GS}$). All data, measured from the same set of devices with a 2.2 nm channel thickness at 10 K, collapse onto a single universal curve, demonstrating the robustness of the scaling behavior. The negative $\beta$ values and a slope close to 1 describe the insulating nature of the film with strong localization. The results reveal that the scaling theory of localization is universal across disordered $In_2O_3$ and other 2D transistors, underscoring its fundamental role in low-dimensional electron transport.

**Conclusion**

In conclusion, we have shown that disorder is a fundamental factor governing charge transport in atomically thin $In_2O_3$ field-effect transistors, inducing a transition from diffusive to localized conduction as carrier density, temperature, and structural order vary.

By systematically tuning film thickness, annealing conditions, and gate bias, we reveal how disorder dictates the breakdown of Ohm's law and the exponential suppression of conductance in the strong localization regime. The extracted localization length quantitatively links microscopic disorder to macroscopic device performance, providing a predictive framework for transport in 2D semiconductors. Most importantly, we achieve the direct experimental realization of the scaling function, validating the scaling theory of localization in a transistor platform. These findings highlight disorder and electron localization as a tunable design parameter for next-generation low-dimensional transistors beyond traditional CMOS scaling.

**Methods**

**Atomic-Layer-Deposition (ALD) and device fabrication.** The device fabrication process began with Si wafers with 90 nm $SiO_2$, followed by a standard cleaning process (ultrasonic rinsing with toluene, acetone, and isopropyl alcohol to remove possible organic particles and dirty materials). Local back gate metal (a 40 nm e-beam evaporated Ni layer) is patterned by photolithography. A 5 nm layer of $HfO_2$ gate dielectric is deposited using ALD at 200 °C. Subsequently, 0.8-3 nm $In_2O_3$ was deposited by ALD at 225 °C using $(CH_3)_3In$ (TMIn) and $H_2O$ as In and O precursors. Film thickness was accurately controlled by the number of ALD cycles. A standard lift-off process (e-beam lithography) was then applied for the 40 nm Ni as source/drain metal. Channel isolation was done by dry etching the $In_2O_3$ layer using Ar plasma. The devices were then annealed at 290 °C in an $O_2$ atmosphere optionally.

**Electrical characterizations.** The fabricated $In_2O_3$ devices were characterized using a Cascade probe station at room temperature in $N_2$ environment at ambient pressure. The temperature-dependent characterization (10 K to 295 K) was performed in a Lakeshore CRX-VF cryogenic probe station. The electrical characterization of transistors were measured with the Keysight B1500A Semiconductor Device Parameter Analyzer.

## Acknowledgements

P.D.Y. was supported by National Science Foundation FuSe2 Program, Semiconductor Research Corporation Global Research Program and Samsung Electronics, Inc.

## Author Contributions

P.D.Y. supervised the project. C.N. designed the experiments. C.N., A.C., J.-Y.L., L.L., Z.L. and Z.Z. fabricated the devices. C.N., A.C., J.-Y.L., and L.L. performed the electrical measurements. C.N. analyzed the data. P.D.Y. and C.N. wrote the manuscript and all the authors commented on it.

## Competing financial interests

The authors declare no competing financial interests.

## Supporting Information

Additional details for parameter scaling effects on transfer characteristics and degree of disorder, large area devices, thickness-, process-, and temperature-dependent electron localization can be found in supplementary information.

**Corresponding Author**

* Peide D. Ye (E-mail: yep@purdue.edu)

# Figures

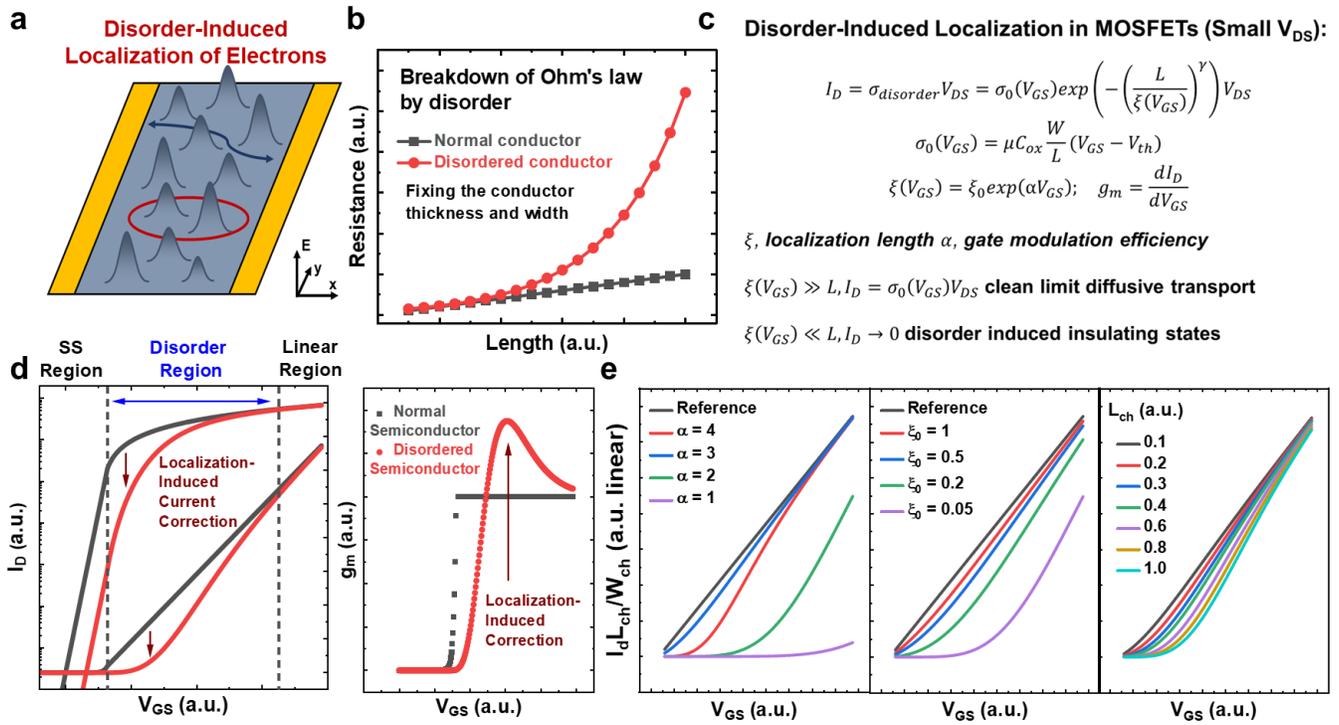

**Figure 1. Effect of disorder-induced electron localization in MOSFETs. a**, Schematic illustration of disorder-induced electron localization, showing the decoherence of electron wavefunctions and their confinement in real space. **b**, Channel-length scaling behavior of a normal (diffusive) conductor and a disordered conductor, highlighting the breakdown of Ohm's law. **c**, Current-disorder relationship derived from the theory of localization. **d**, Conceptual comparison of transfer characteristics and $g_m$ between a conventional Si-based FET and a disordered oxide-semiconductor FET. **e**, Illustration of the interplay among channel length, localization length, and gate-modulation efficiency, and their combined influence on the transfer characteristics.

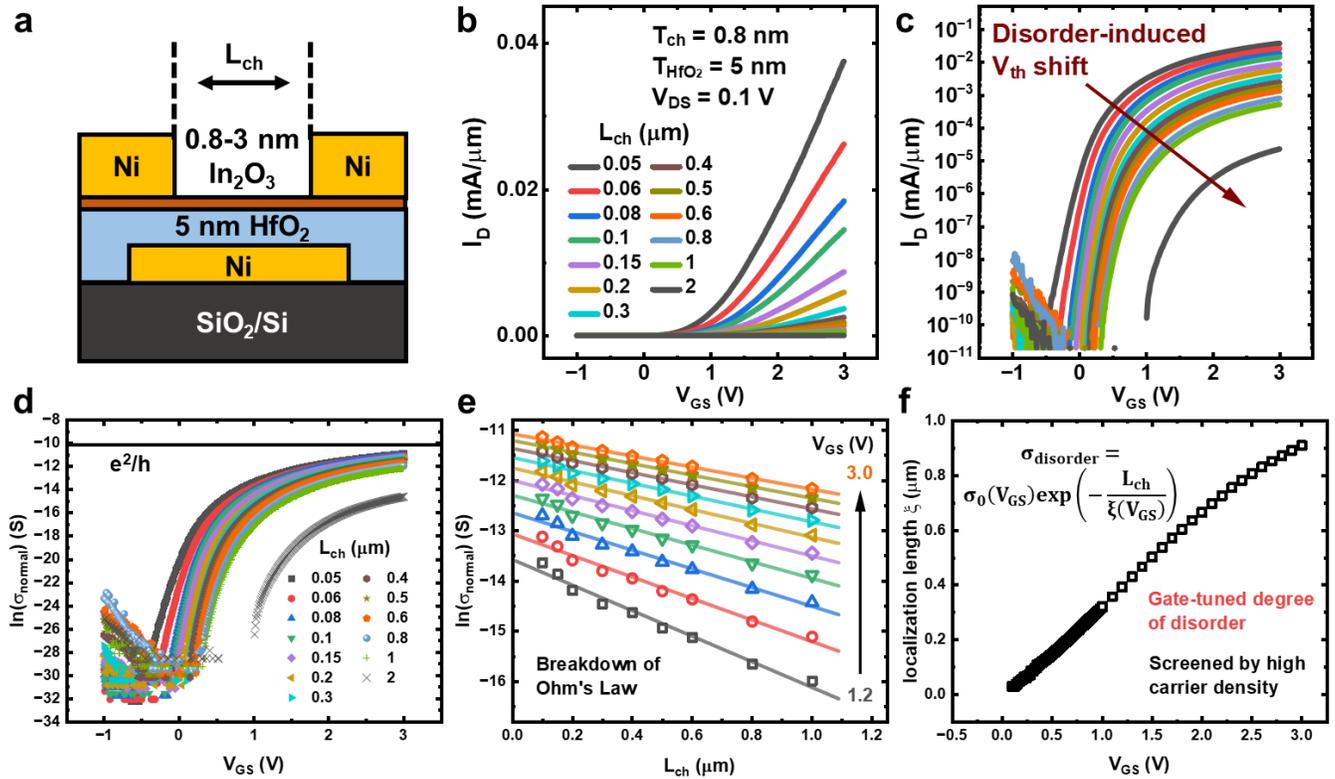

**Figure 2. Electron localization in In$_2$O$_3$ FETs. a**, Schematic of the In$_2$O$_3$ FET device structure used to investigate disorder-induced electron localization. **b-c**, Channel-length-dependent transfer characteristics of In$_2$O$_3$ FETs with a 0.8 nm channel thickness, plotted on linear (**b**) and logarithmic (**c**) scales. The V$_{th}$ shift with increasing channel length arises from localization effects. **d**, Normalized conductivity (σ/L$_{ch}$) as a function of gate voltage for devices with different channel lengths, showing the breakdown of Ohm's law. **e**, Channel-length-dependent conductivity scaling exhibiting an exponential decrease at various carrier densities (gate voltages). **f**, Localization length extracted from the slopes in **e** as a function of gate voltage. The significant difference of 2 μm data from the rest of the group data in (**c**) and (**d**) highlight the profound localization effect or breakdown of linear scaling.

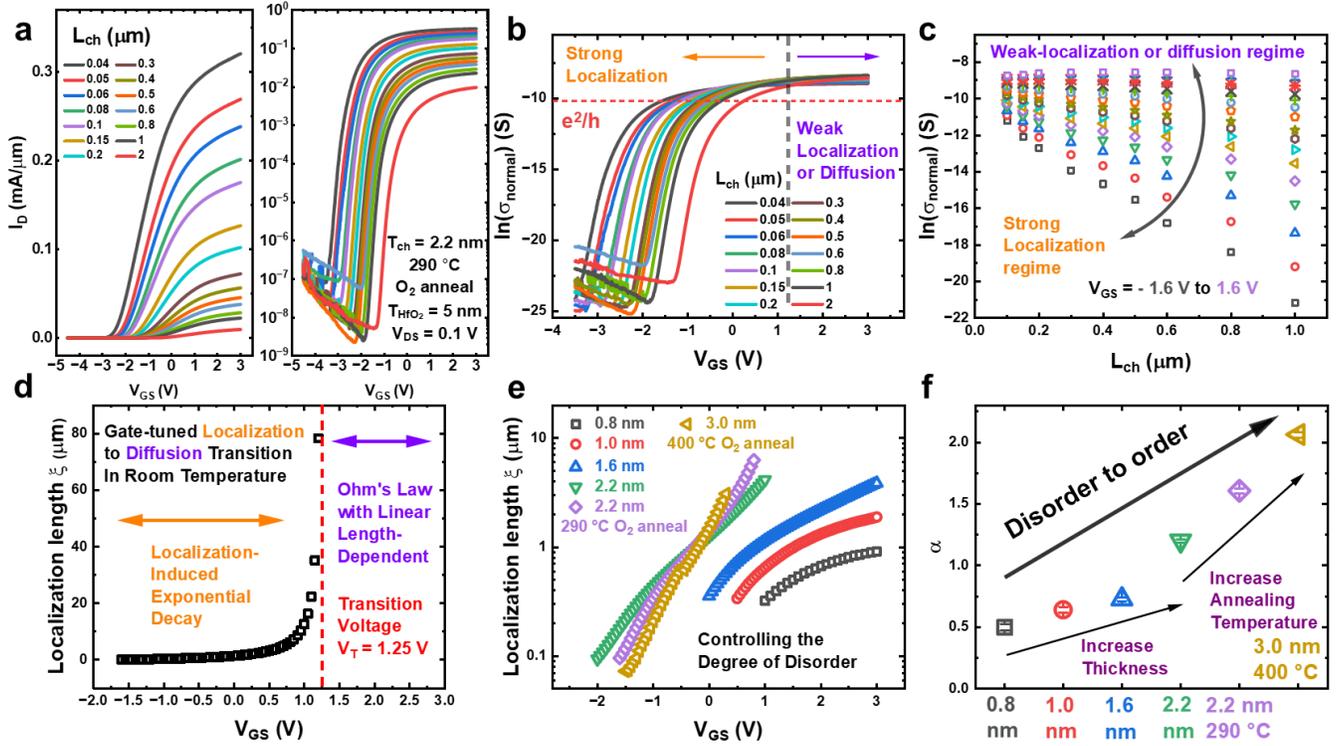

**Figure 3. Gate-tunable diffusion-localization transition and the role of thickness and annealing temperature. a**, Transfer characteristics of $In_2O_3$ FETs with a 2.2 nm channel thickness after post $O_2$ annealing at 290 °C, shown on both linear and logarithmic scales. **b**, Gate-voltage dependence of the normalized conductivity for different channel lengths, illustrating the gate-tunable transition from diffusive to localized transport. **c**, Channel-length-dependent conductivity scaling exhibiting an exponential decrease, signifying the breakdown of Ohm's law. **d**, Gate-voltage-dependent localization length showing convergence at certain gate bias. **e**, Localization length as a function of gate voltage under different channel thicknesses and annealing temperatures. **f**, Parameter α extracted from **e**, showing that increasing film thickness and annealing temperature effectively reduces the degree of disorder.

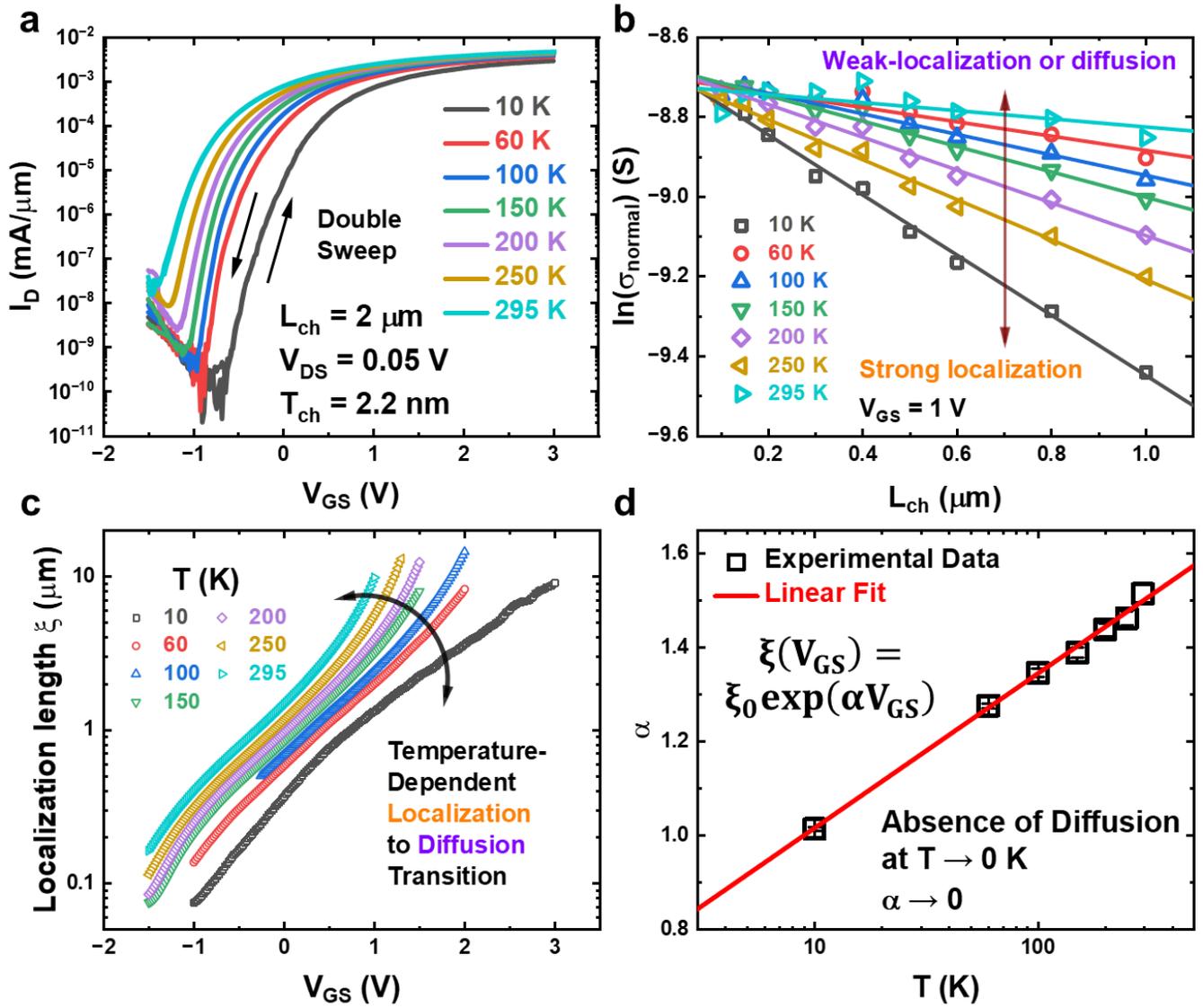

**Figure 4. Temperature dependence of diffusion-localization transition. a,** Temperature-dependent transfer characteristics of an $In_2O_3$ FET with a 2 μm channel length measured from 295 K to 10 K. The forward and reverse gate-voltage sweeps show negligible hysteresis. **b,** Temperature-dependent conductivity scaling curves at $V_{GS}$ = 1 V, illustrating the transition from diffusive to localized transport with decreasing temperature. **c,** Gate-voltage-dependent localization length extracted at different temperatures from the same set of devices. **d,** Temperature dependence of the parameter α obtained from the slope

in **c**. The linear trend of α toward zero indicates the absence of diffusion and gate tunability as T → 0 K, a characteristic signature of low-dimensional electronic systems.

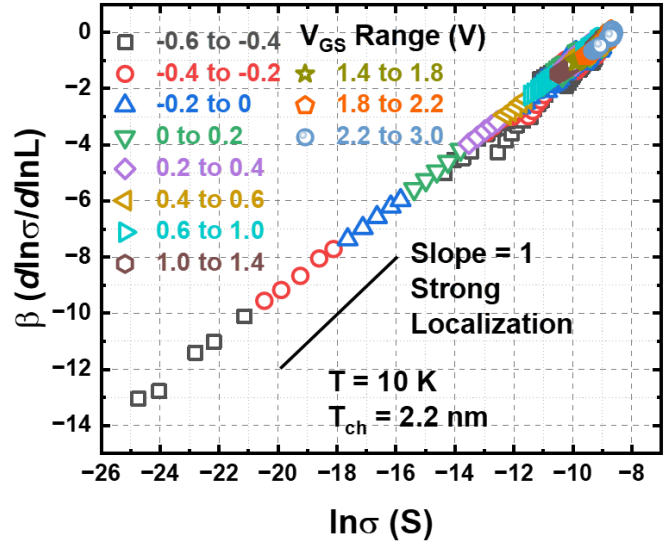

**Figure 5. Scaling theory of electron localization and disorder.** Experimental extraction of the scaling theory of localization, showing the relationship between conductivity σ and the scaling factor β = $d\ln\sigma/d\ln L$. By tuning the gate voltage $V_{GS}$, different conductivities (carrier densities) are accessed within the same device. The extracted data at various carrier densities collapse onto a single line with a slope close to 1, characteristic of the strong-localization regime. The negative values of β indicate localization-dominated transport in the disordered film, and the observed scaling behavior offers practical guidelines for understanding and designing low-dimensional electronic systems where disorder plays a key role.

Supplementary Information for:

# Breakdown of Ohm's Law by Disorders in Low-Dimensional Transistors

Chang Niu, Adam Charnas, Jian-Yu Lin, Linjia Long, Zehao Lin, Zhuocheng Zhang and Peide D. Ye*

*Elmore Family School of Electrical and Computer Engineering and Birck Nanotechnology Center, Purdue University, West Lafayette, IN 47907, United States.*

*Correspondence and requests for materials should be addressed to P. D. Y. (yep@purdue.edu)

# List of contents:

**Supplementary figures:**

**Supplementary notes:**

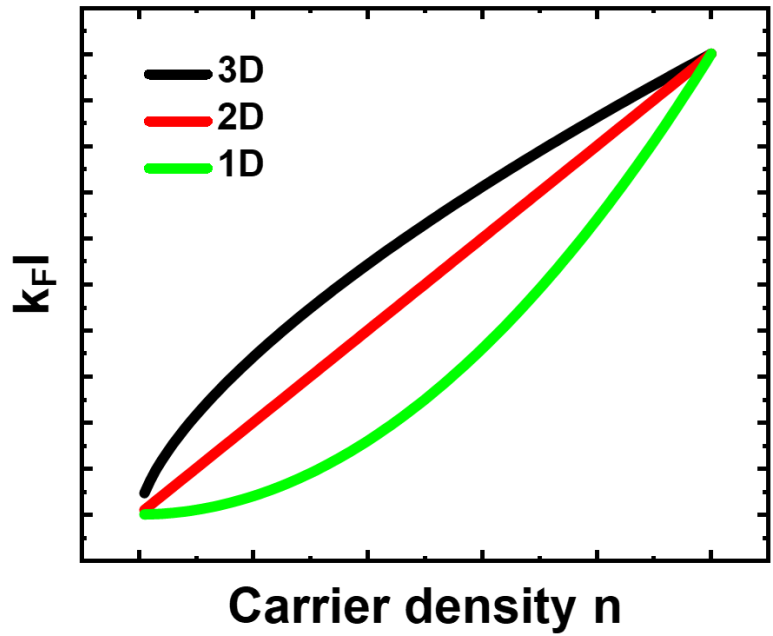

**Figure S1. Degree of disorder scaling with carrier density in different dimensions.** With decreasing dimensionality, the degree of disorder scales more rapidly with carrier density, indicating that disorder effects become increasingly pronounced and influential in lower-dimensional transistors.

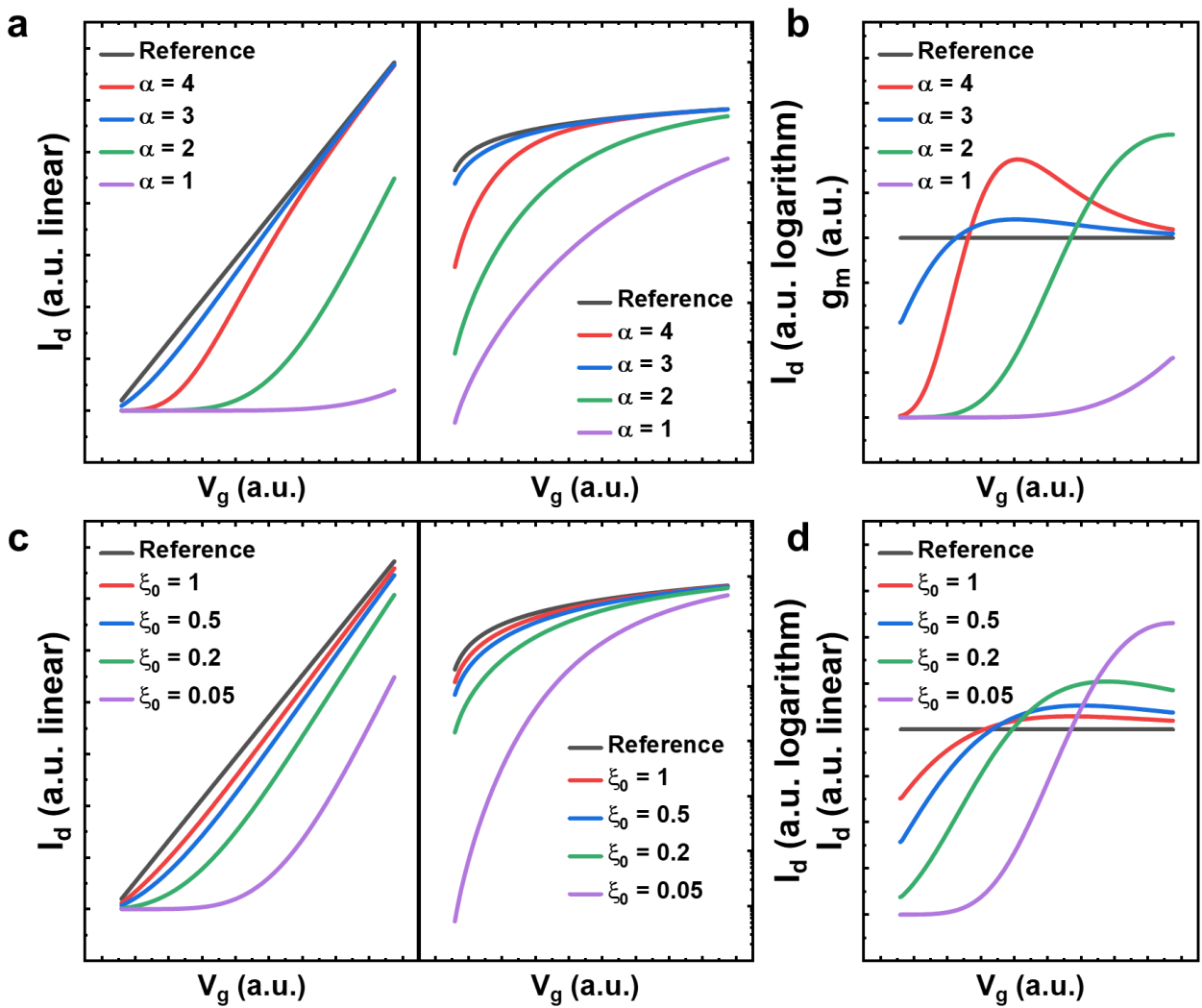

**Figure S2. Transfer characteristics and transconductance at different parameters. a**, Transfer characteristics plotted on linear and logarithmic scales for different values of α, showing that faster gate modulation of the localization length yields device characteristics approaching ideal performance. **b**, corresponding $g_m$ curves at different α. **c**, Transfer characteristics plotted on linear and logarithmic scales for different initial localization lengths ($\xi_0$), indicating that larger $\xi_0$ results in improved device performance closer to the ideal case. **d**, Corresponding $g_m$ plot at different $\xi_0$.

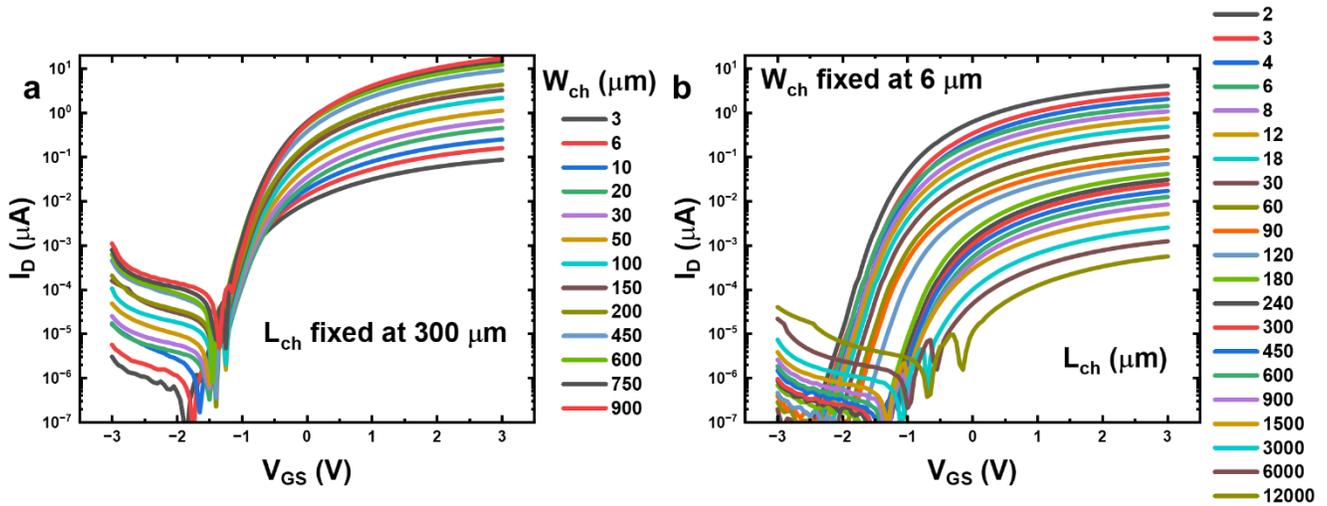

**Figure S3. Electron localization in large millimeter-scale devices. a**, Channel-width-dependent transfer characteristics with $W_{ch}$ ranging from 3 μm to 900 μm. The small threshold-voltage shift indicates that channel width plays a less significant role than channel length in electron localization. **b**, Channel-length-dependent transfer characteristics with $L_{ch}$ ranging from 2 μm to 12 000 μm. The similar $V_{th}$ shift observed across all lengths, consistent with the main-text discussion, confirms that localization originates from intrinsic channel material properties. The influence of contact resistance is negligible in these long-channel devices.

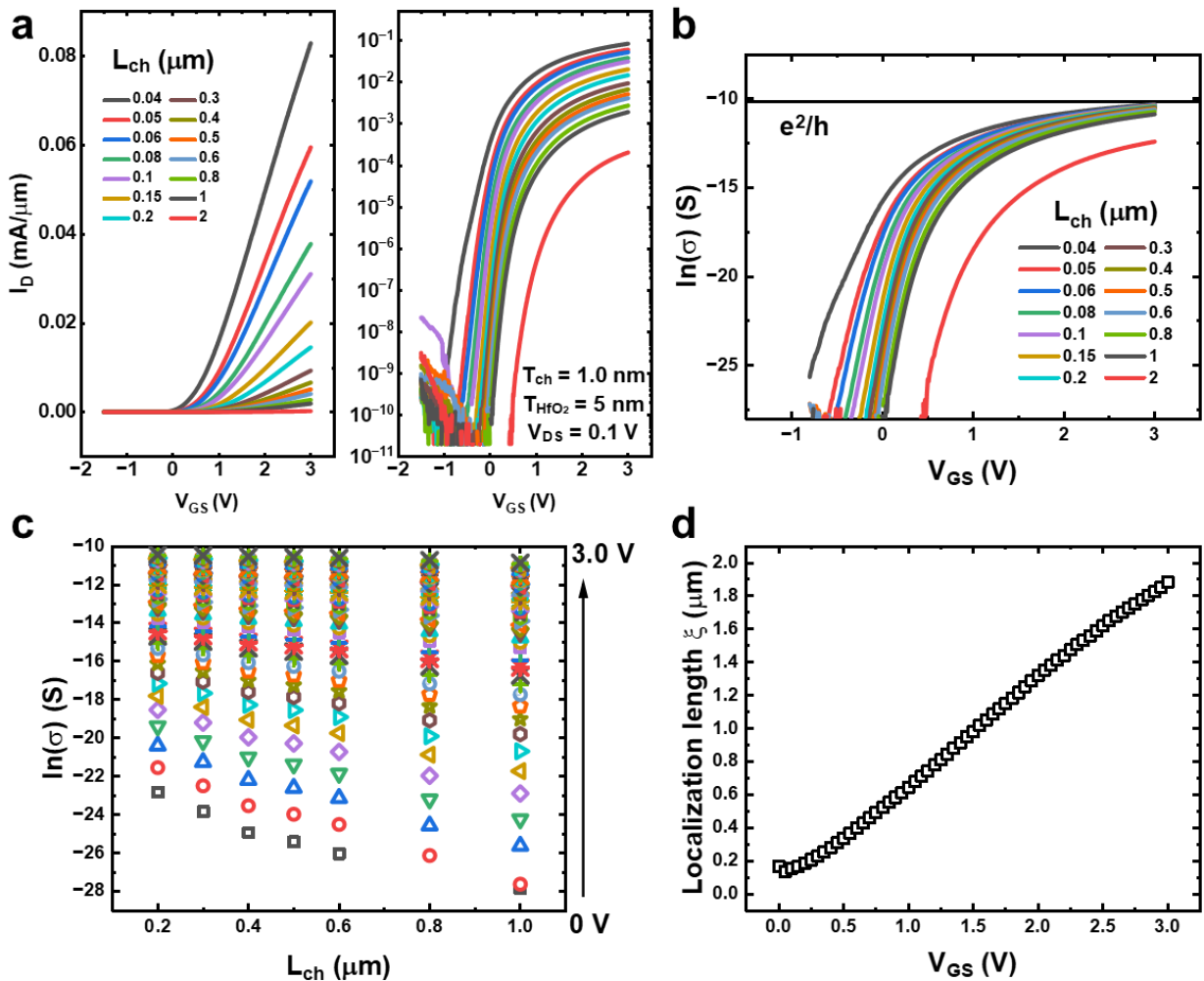

**Figure S4. Electron localization in 1.0 nm thick In$_2$O$_3$. a**, Transfer characteristics of In$_2$O$_3$ FETs with a 1.0 nm channel thickness with no post O$_2$ annealing, shown on both linear and logarithmic scales. **b**, Gate-voltage dependence of the normalized conductivity for different channel lengths. **c**, Channel-length-dependent conductivity scaling exhibiting an exponential decrease. **d**, Gate-voltage-dependent localization length extracted from the data in **c**.

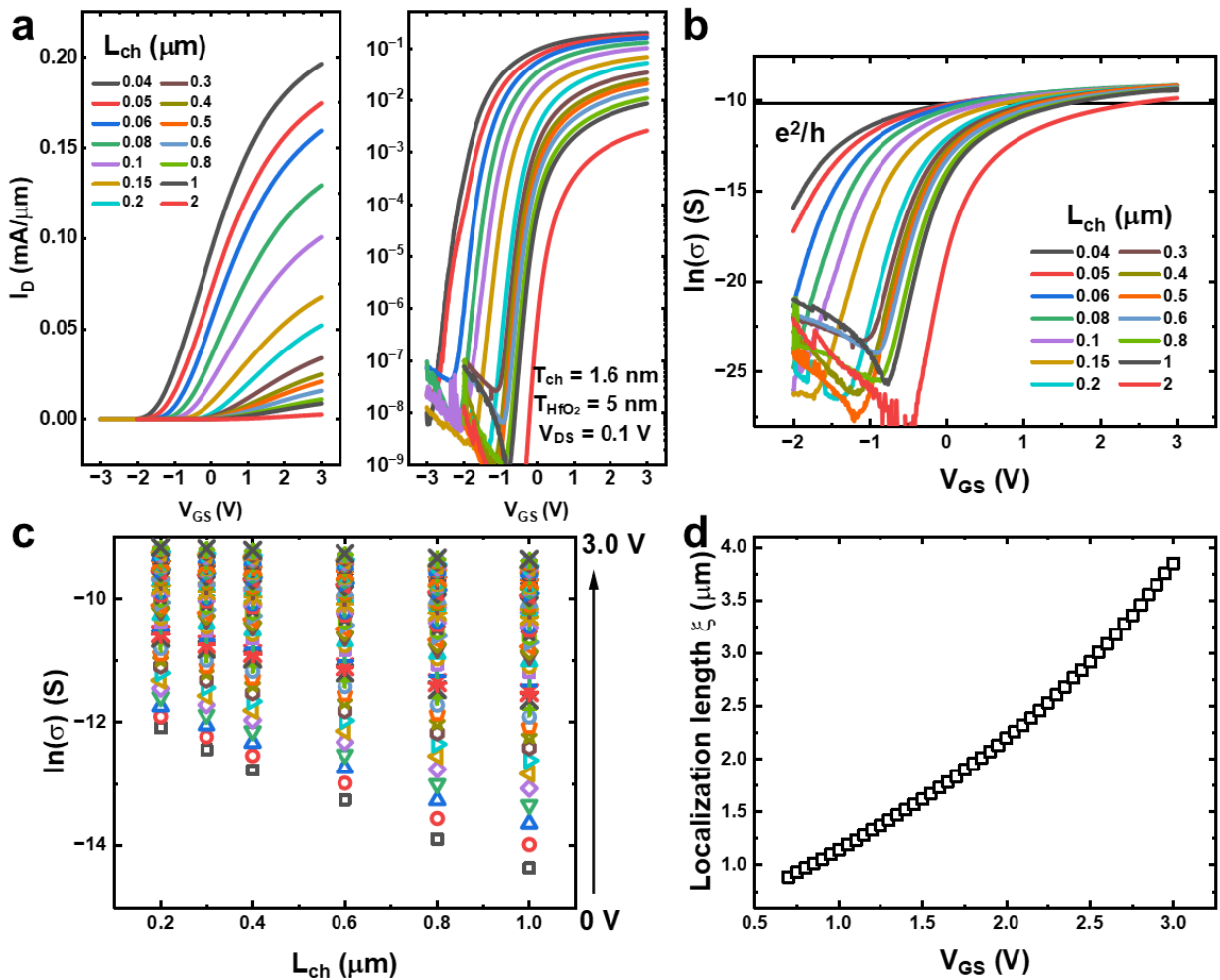

**Figure S5. Electron localization in 1.6 nm thick $In_2O_3$. a**, Transfer characteristics of $In_2O_3$ FETs with a 1.6 nm channel thickness with no post $O_2$ annealing, shown on both linear and logarithmic scales. **b**, Gate-voltage dependence of the normalized conductivity for different channel lengths. **c**, Channel-length-dependent conductivity scaling exhibiting an exponential decrease. **d**, Gate-voltage-dependent localization length extracted from the data in **c**.

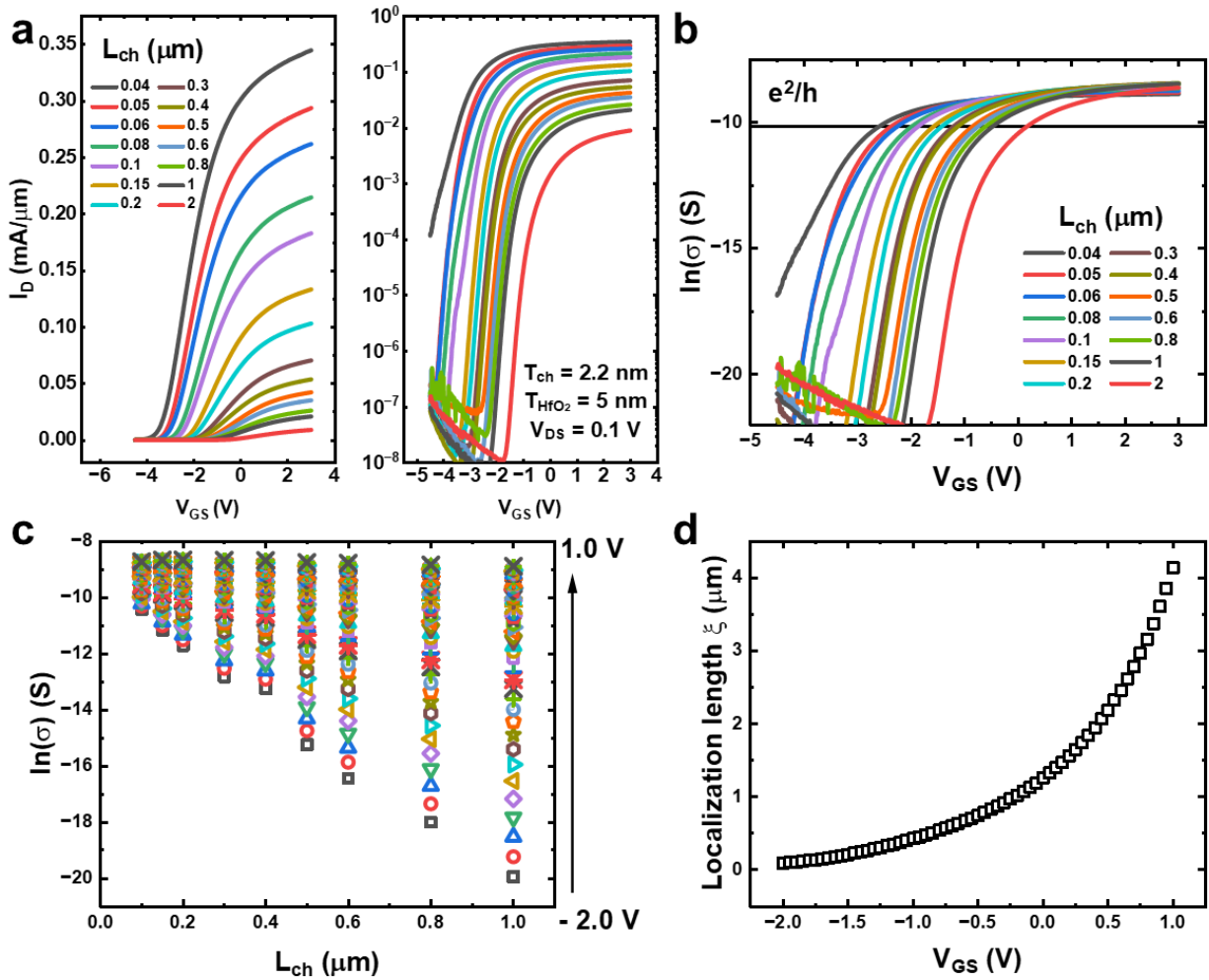

**Figure S6. Electron localization in 2.2 nm thick In$_2$O$_3$. a**, Transfer characteristics of In$_2$O$_3$ FETs with a 2.2 nm channel thickness with no post O$_2$ annealing, shown on both linear and logarithmic scales. **b**, Gate-voltage dependence of the normalized conductivity for different channel lengths. **c**, Channel-length-dependent conductivity scaling exhibiting an exponential decrease. **d**, Gate-voltage-dependent localization length extracted from the data in **c**.

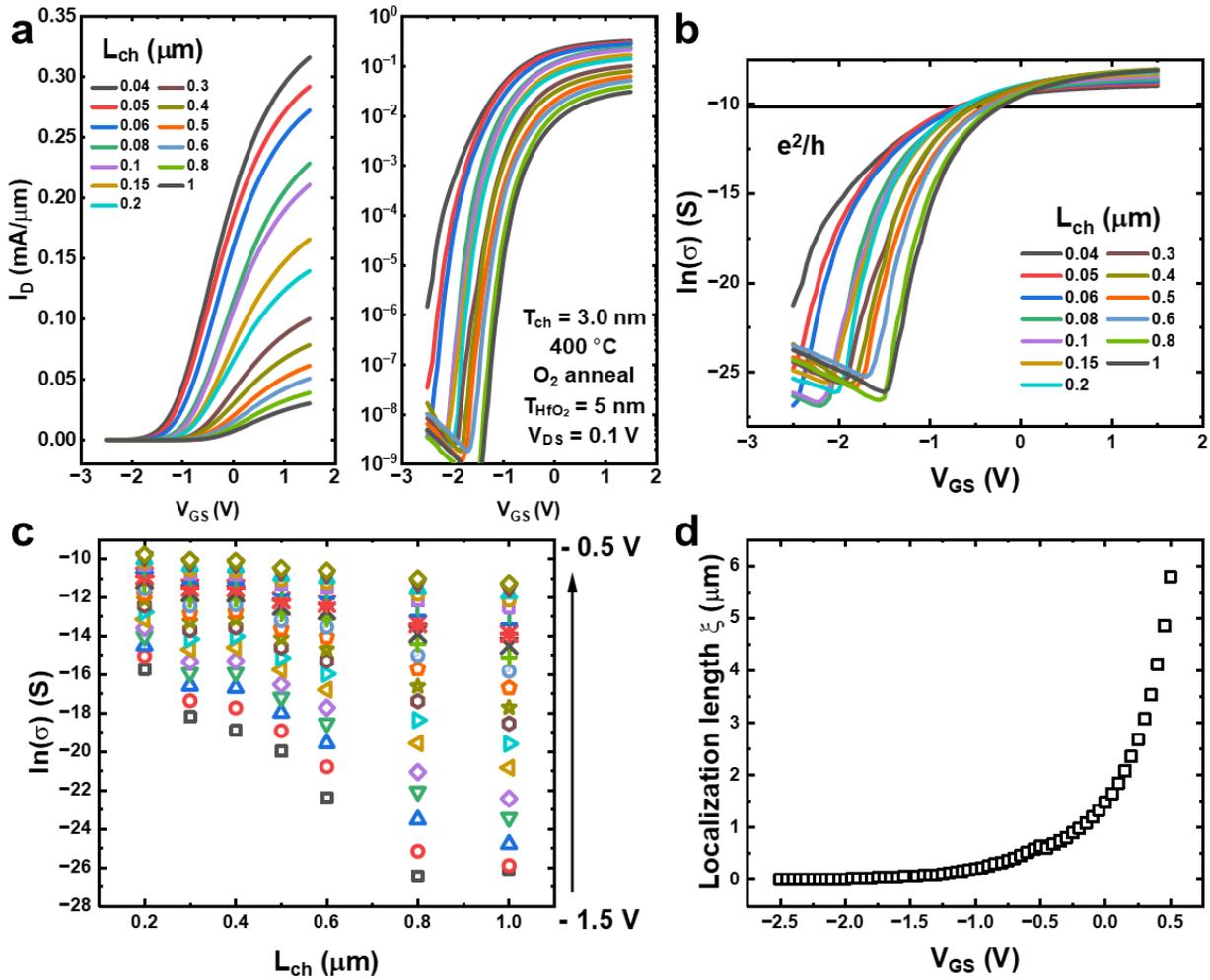

**Figure S7. Electron localization in 3.0 nm thick $In_2O_3$ with 400 °C $O_2$ annealing. a**, Transfer characteristics of $In_2O_3$ FETs with a 3.0 nm channel thickness with post $O_2$ annealing at 400 °C, shown on both linear and logarithmic scales. **b**, Gate-voltage dependence of the normalized conductivity for different channel lengths. **c**, Channel-length-dependent conductivity scaling exhibiting an exponential decrease. **d**, Gate-voltage-dependent localization length extracted from the data in **c**.

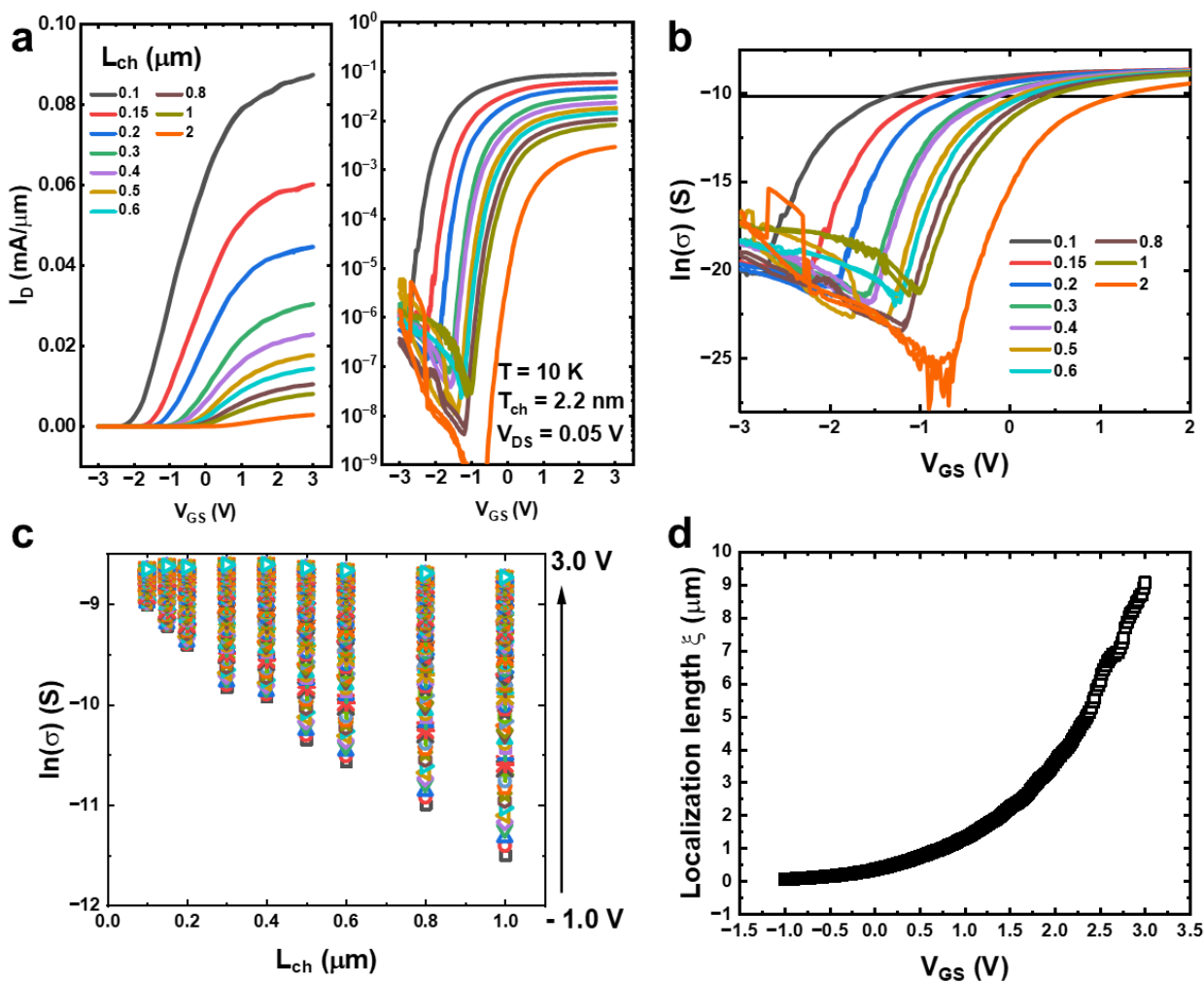

**Figure S8. Electron localization in 2.2 nm thick In$_2$O$_3$ at 10 K. a**, Transfer characteristics of In$_2$O$_3$ FETs with a 2.2 nm channel thickness with no post O$_2$ annealing measured at 10 K, shown on both linear and logarithmic scales. **b**, Gate-voltage dependence of the normalized conductivity for different channel lengths. **c**, Channel-length-dependent conductivity scaling exhibiting an exponential decrease. **d**, Gate-voltage-dependent localization length extracted from the data in **c**.

**Supplementary notes:**

**Note S1: Degree of disorder in different dimensions.**

The mean free path is:

$$l = v_F \tau$$

where $v_F = \hbar k_F / m^*$ is the Fermi velocity and $\tau$ is the relaxation time.

Mobility is:

$$\mu = \frac{e\tau}{m^*}$$

Combine them:

$$l = \frac{\hbar k_F \mu}{e}$$

Carrier density in different dimensions:

$$n_{1D} = \frac{g_s g_v k_F}{\pi}, \quad n_{2D} = \frac{g_s g_v k_F^2}{4\pi}, \quad n_{3D} = \frac{g_s g_v k_F^3}{6\pi^2}$$

Substitute into $k_F l$:

$$k_F l = \frac{\hbar \mu}{e}\left(\frac{\pi n_{1D}}{g_s g_v}\right)^2, \quad \frac{\hbar \mu}{e}\left(\frac{4\pi n_{2D}}{g_s g_v}\right) = \frac{4\pi \hbar}{e^2}\sigma_\square, \quad \frac{\hbar \mu}{e}\left(\frac{6\pi^2 n_{3D}}{g_s g_v}\right)^{\frac{2}{3}}$$

where $\sigma_\square = e n_{2D} \mu$